\documentstyle[aps,epsfig,multicol]{revtex}

\newcommand{\bq}{\begin{equation}}
\newcommand{\eequ}{\end{equation}}
\newcommand{\bqa}{\begin{eqnarray}}
\newcommand{\eqa}{\end{eqnarray}}
\newcommand{\nn}{\nonumber}
\newcommand{\ms}[1]{\mbox{\scriptsize #1}}

\newcommand{\smallfrac}[2]{\mbox{$\frac{#1}{#2}$}}

\newcommand{\half}{\smallfrac{1}{2}}

\begin{document}
\draft

\title{Information, disturbance and Hamiltonian quantum feedback control                 \vbox to 0pt{\vss
                    \hbox to 0pt{\hskip-37pt\rm LA-UR-00-1937\hss}
                    \vskip 25pt}}

\author{Andrew C. Doherty$^{\star}$, Kurt Jacobs$^{\dagger}$, Gerard Jungman$^{\dagger}$}
\address{$^{\star}$ Norman Bridge Laboratory of Physics 12-33,
California Institute of Technology, Pasadena CA 91125}
\address{$^{\dagger}$ T-8, Theoretical Division, Los Alamos National
  Laboratory, Los Alamos, New Mexico 87545.}

\maketitle

\begin{abstract}
  We consider separating the problem of designing Hamiltonian quantum
  feedback control algorithms into a measurement (estimation) strategy
  and a feedback (control) strategy, and consider optimizing
  desirable properties of each under the minimal constraint that the
  available strength of both is limited. This motivates concepts of
  information extraction and disturbance which are distinct from those
  usually considered in quantum information theory. Using these 
  concepts we identify an information trade-off in quantum feedback 
  control.
\end{abstract}

\begin{multicols}{2}

\section{Introduction}
With experimental advances, particularly in the fields of cavity QED~\cite{CQED} and ion trapping~\cite{ion}, it is possible to observe individual quantum systems in real time, and it is therefore natural to consider the possibility of controlling such systems in real time using feedback~\cite{Bel1}. As a special case, real-time Markovian quantum feedback has been analyzed~\cite{qfb1,sloss,dunn,hof} and implemented experimentally in certain quantum optical systems~\cite{taub}. Feedback control is invaluable in macroscopic applications, and as a consequence there is a vast body of literature devoted to classical feedback control.  While results from classical control theory may be applied fruitfully to the quantum domain in certain special cases~\cite{BelLQG,DJ}, these are not adequate in general, primarily because quantum measurement is quite different in nature from classical measurement, in that it has the capacity to disturb the system under observation~\cite{DHJMT}. As a result, the development of optimal quantum control strategies requires optimizing over possible measurement strategies, which is unnecessary in classical control.

In feedback control the dynamics of a system is manipulated by using information obtained about the system through measurement. The goal is usually to maintain a desired state or dynamics in the presence of noise. A central problem of feedback control theory is the development of algorithms to achieve this goal. The approach to controller design that we consider here is to examine the measurement and feedback steps separately, thereby splitting the feedback control problem into two parts. One can then consider optimizing desirable properties of these parts separately under suitable constraints. If one allows the strength of either measurement or Hamiltonian feedback to be infinite, then any control objective can be achieved perfectly (this will be shown below once we have made these concepts of strength more precise). A constraint on strength is therefore the minimal constraint under which the problem of quantum feedback control is non-trivial, and this is the constraint we employ here.

The action of optimizing for the feedback and measurement independently ignores the possibility that truly optimal solutions may require considering both together. We will also simplify the problem by considering the optimization at each time step separately. This assumes that it is never desirable to perform worse at the current time in order to perform better at some future time. The approach we take here is therefore not aimed at finding a globally optimal solution given a set of constraints. However, the expectation is that the concepts we introduce here provide a simple systematic approach which one can expect to produce good results, and provide an insight into the kind of measurement processes which are desirable in feedback control.

For the feedback step, we consider the question of the effectiveness of the control by defining a cost function. Since one is interested in controlling the dynamics of a given quantum system (usually in the presence of some unavoidable source of environmental noise), one can specify the objective by specifying the most desired state for the system at each instant. The `cost' function is then the sum of the distances of the state of the system from the desired state at each point in time, for some suitable measure of distance. Finding the optimal control strategy then consists in minimizing the cost function, under suitable constraints for the strength of the feedback. Note that this is somewhat different from the standard approach taken in modern classical control theory~\cite{Jacobs,Maybeck,Whittle}, and more similar to the approach taken in the new techniques of `postmodern' classical control~\cite{Doyle}. In modern classical control (e.g. LQG theory) one usually optimizes a `total' cost function obtained from a suitably weighted sum of the cost function defined here, and another cost function intended to capture the cost of feedback strength. 


We will restrict ourselves to control objectives such that the desired state at each time (the target state) is pure. Impurity (mixing) merely signifies that our knowledge of the quantum system is less than maximal, which is by assumption undesirable. 


In considering the optimality of the measurement step, rather than attempting to find a measurement which explicitly optimizes the cost function, we define concepts of information and disturbance, motivated by the feedback control problem. We then consider finding measurements which maximize the information and minimize the disturbance. We find that in general these two targets are mutually exclusive, in striking contrast to classical control theory. This implies the existence of a trade-off between information and disturbance in quantum feedback control. 

Since we focus on continuous feedback control, and many readers will be familiar with generalized measurements but unfamiliar with the formalism of continuous quantum measurement, we describe in the next section how continuous observation is formulated within the language of generalized measurements. In Section~\ref{sec:ms} we define the concept of the strength of a measurement, required as a minimal constraint for the feedback control problem. In Section~\ref{MandF} we discuss in detail the division of feedback control into `pure' measurement and Hamiltonian feedback, and consider what may be achieved when there is no limitation on the strength of either. We also discuss what may be achieved in this case both without feedback and with measurement-only feedback. In Section~\ref{maxmin}, we consider the measurement process, define
concepts of information and disturbance, and consider minimizing the disturbance and maximizing the information. In Section~\ref{opham} we examine the Hamiltonian feedback and obtain Hamiltonians which minimize the instantaneous cost function. In Section~\ref{sim} we implement the feedback control of a two-state system, showing how the ideas presented in the previous sections manifest in the performance of the control algorithm. Section~\ref{conc} summarizes and concludes.

\section{Continuous observation and generalized measurements}
\label{contm}
We will concern ourselves primarily with continuous-time quantum
feedback control, in which a system is observed continuously, and the
results of the measurements (the {\em measurement record}) used to
continuously alter the Hamiltonian of the system to effect control.
We now discuss how continuous observation may be described within the
language of generalized quantum measurements, implemented as positive operator
valued measures (POVM's).

Continuous measurements on a quantum system generate a measurement
record that is a continuous-time stochastic process, which may be
either a (Gaussian) Wiener process or a Poisson process~\cite{Traj,BelLQG}. 
For a given physical system, these two kinds of processes will result 
from making different measurements, for example photon counting and 
homodyne detection of optical beams.

The key ingredient in describing continuous measurements is that
during an infinitesimal time step $dt$, the information obtained by
the observer must scale as $dt$, so that one can take the continuum
limit and obtain a sensible answer. This may be realized by defining a
POVM, given by $\int \Omega_\alpha^\dagger \Omega_\alpha d\alpha=1$,  to
describe the result of an observation in the time interval $dt$ by~\cite{MC}
\begin{equation}
   \Omega_{\alpha} = \left( \frac{\pi}{2dt}\right)^{1/4} e^{-kdt(Q-\alpha)^2} ,
\end{equation}
where $Q$ is a arbitrary operator for the system under observation,
$\alpha$ takes all values on the real line, and $k$ is a positive real
constant. For reasons that will be made clear in the next section, we
will only need to be concerned with the case in which $Q$ is
Hermitian, so that $Q$ may be referred to as an observable, and we
will assume this in what follows. Note that each $\Omega_\alpha$ is 
a weighted sum
of projectors onto the eigenbasis of $Q$, where the weighting is
peaked at $\alpha$. Thus each application of the one of the $\Omega's$
provides some information about the observable $Q$. However, as $dt$
tends to zero, this information also tends to zero, since the
$\Omega's$ become increasingly broad over the eigenstates of $Q$.
Calculating the measurement result in the interval $dt$ at 
time $t$, and denoting this as $dy(t)$, we have
\begin{equation}
   dy(t) = 4k\langle Q\rangle dt + \sqrt{2k}dW ,
\label{mrec}
\end{equation}
where $dW$ is the Wiener increment for the interval $dt$. Using this,
one can obtain the stochastic evolution of the quantum state under this
measurement process, referred to as a quantum trajectory, and this is
given by the Stochastic Master Equation (SME)
\begin{eqnarray}
  d\rho & = & -i[H,\rho]dt - k[Q,[Q,\rho]]dt \nn \\
        &   & + (Q\rho + \rho Q - 2\mbox{Tr}[Q\rho]\rho)\sqrt{2k} dW .
\label{SME}
\end{eqnarray}
where $H$ gives the system evolution in the absence of the measurement.
We can also readily obtain the non-selective evolution, in which the
measurement results are ignored, and this is given by
\begin{eqnarray}
   \rho(t+dt) & = & -i[H,\rho]dt + \int \Omega_\alpha \rho \Omega_\alpha^\dagger d\alpha \nn \\
              & = &  -i[H,\rho]dt -k[Q,[Q,\rho(t)]]dt .
\end{eqnarray}
When $H$ commutes with $Q$ this evolution leads to a diagonalization of 
$\rho$ in the basis of
$Q$, as one would expect for a measurement of $Q$. Similarly,
integrating the SME in this case, one finds that the result in the long 
time limit
is a projection onto one of the eigenstates of $Q$. Such a POVM
realizes a continuous measurement of the operator $Q$, such that the
measurement record is a Wiener process.

One can also define a POVM to provide continuous observation in which
the measurement record is a Poisson process. Since this requires only
one of the two possible outcomes at each interval $dt$, the POVM
consists of only two measurement operators:
\begin{eqnarray}
  \Omega_0 & = & 1 - \half k Q^2 dt , \\
  \Omega_1 & = & Q \sqrt{k dt} .
\end{eqnarray}
That this gives a Poisson process can be seen by considering the
probabilities for the outcomes 0 and 1, which are $1-k\langle Q\rangle
dt$ and $k\langle Q\rangle dt$, respectively. Result 1 therefore
corresponds to an Poisson `event', which happens occasionally, and 0 to
the absence of one. The SME corresponding to the measurement process
is different from that corresponding to the Wiener measurement, but
the non-selective evolution is identical. Physically, the non-selective
evolution is fixed by choosing the interaction of the system with the
environment which is mediating the measurement, and the trajectory,
whether Poisson or Wiener, is selected by how one chooses to measure
the environment so as to extract the information about the system. In
fact, by taking a suitable unitary transformation of the Poisson
measurement operators, and taking the appropriate limit in which there
are many events in each interval $dt$, one can obtain the Wiener process
measurement from the Poisson measurement, and so the first can be
regarded as a special case of the second. This is discussed in detail
in~\cite{HMWPhD,KJPhD}.

The point we wish to note here is that regardless of how one chooses
the trajectory, a continuous measurement of an observable $Q$ is given
by a POVM in which all the measurement operators $\Omega_\alpha$ are
positive operators, diagonal in the basis of $Q$, and one must merely
be careful to choose the form of these operators with respect to $dt$
so as to provide a sensible continuum limit.

\section{The Strength of a Measurement}~\label{sec:ms}
Clearly the more accurate the measurements of the observer, the more
information she is able to obtain, and better able she is to choose 
feedback to effectively control the system. However, in general, more
accurate measurements require more resources. In treating quantum 
feedback control, it is sensible to consider a restriction on available 
resources, and hence a restriction on measurement accuracy. To treat this 
quantitatively, one must introduce a sufficiently precise notion of the 
accuracy, or {\em strength}, of a quantum measurement. 

For the purposes of feedback control, since it is the final state 
resulting from measurement that the observer must act upon with 
feedback, it is the observers information about this {\em final} state 
which is relevant. Intuitively, one can therefore think of stronger 
measurements as providing, on average, final states which are 
more pure (or, alternatively, have a smaller von Neumann entropy) than 
weaker measurements. When considering continuous observation, in the 
absence of any noise sources, an initially impure state is continually 
purified. In this case the strength of the measurement can be thought 
of as being proportional to the {\em rate} of this purification. 
Note that this concept of information extraction by a measurement is 
quite different from that usually considered in quantum 
information theory. There, authors have been concerned about
the information that a measurement provides about the initial state of 
the system (the state immediately before the measurement)
~\cite{Massar,Derka}, whereas in our case it is the information about 
the final state which is important.

We will not need an explicit definition for measurement strength here, 
since we will only require two properties of measurement strength which we 
will motivate below. However, we will give an example of an explicit 
definition which satisfies these two properties. To motivate the first 
property, we note that as we have defined it so far, it is clear that 
the strength of a measurement in some sense characterizes the average 
rank of the operators $\Omega_m$ which make up the associated POVM 
($\sum_m \Omega_m^\dagger \Omega_m = 1$). If all the $\Omega_m$ are rank 
one, then one always obtains a pure final state, and therefore complete 
information, regardless of the initial state. The higher the rank of the 
projectors, the higher in general will be the von Neumann entropy for a 
fixed initial state. The first property we will require is that measurements 
that consist of rank one projectors should have maximum strength (for 
measurements on a system of a given dimension). 

For the remainder of this paper, we will refer to measurements 
for which at least one of the $\Omega_m$ are rank one as infinite strength measurements. This terminology is natural in the context of continuous 
observation, since in order to provide rank one projections in a finite 
time from a continuous measurement, one would have to take the limit 
$k\rightarrow\infty$ in Eq.(\ref{SME}). 
However, we wish to stress that our use of this terminology is not 
intended to imply that any explicit definition of measurement strength 
should necessarily take this value for these kinds of measurements.

The second property we wish to impose is that strength be invariant under
unitary transformations of the measurement operators. To motivate this 
property, one can consider a device which measures the spin of a two state 
system. One would expect such a device to provide the same strength of 
measurement regardless of how it is oriented in space. Since spatial 
rotation covers all unitary transformations for a spin-half, for this 
system strength should be invariant under all unitary transformations 
of the $\Omega_n$. We will explicitly consider the spin-half system later.

To provide an example of an explicit definition of measurement strength 
for single-shot measurements on finite dimensional systems, one can first 
consider the average uncertainty after the measurement result is known. 
Using the von Neumann entropy, for a measurement described by $\sum_n 
\Omega_n^\dagger\Omega_n = 1$, this is
\begin{equation}
  u_{\ms{V}}(\rho) = \sum_n \mbox{Tr}[\Omega_n \rho \Omega_n^\dagger
       \mbox{ln}(\Omega_n \rho \Omega_n^\dagger/\mbox{Tr}[\Omega_n \rho
       \Omega_n^\dagger])],
\end{equation}
where $\rho$ is the initial state of the system. Using the purity as an 
alternative measure of uncertainty we have
\begin{equation}
  u_{\ms{p}}(\rho) = 1 - \sum_n\frac{\mbox{Tr}[(\Omega_n \rho \Omega_n^\dagger)^2]}{\mbox{Tr}[\Omega_n\rho \Omega_n^\dagger]}.
\end{equation} 
To obtain definitions of measurement strength that 
satisfy our two properties we can use the following functions of these 
uncertainties:
 \begin{eqnarray}
    s_{\ms{v}} & = & \frac{1}{ u_{\ms{V}}(I/N) } - \frac{1}{\log(N)} ,   
    \label{sdef1} \\
    s_{\ms{p}} & = & \frac{1}{ u_{\ms{p}}(I/N) } - \frac{N}{(N-1)} , 
    \label{sdef2}
\end{eqnarray}
in which $N$ is the dimension of the system being measured. These 
particular definitions have the additional property that they tend to  
infinity for von Neumann measurements, but no others. It is also a fairly 
simple matter to write explicit definitions of measurement strength that 
tend to infinity for measurements that contain at least one rank one 
projector.
 
Definitions of measurement strength for single-shot measurements may be 
extended to continuous measurements by using the 
{\em rate} of uncertainty reduction. Using the explicit definitions given 
above (Eq.(\ref{sdef1}) and Eq.(\ref{sdef2})), it is straightforward to 
calculate this rate from Eq.(\ref{SME}) and the Ito rules for stochastic 
differential equations~\cite{Gardiner1}:
\begin{eqnarray}
    \frac{d}{dt}s_{\ms{v}} & = & \frac{4k}{N \log(N)^2} \left( \mbox{Tr}[Q^2]+3\mbox{Tr}[Q]^2 \right) \\
   \frac{d}{dt}s_{\ms{p}} & = & 8k\frac{N^2}{ (N-1)^2}\mbox{Tr}[Q^2].
\end{eqnarray}

In considering feedback control, measurement strength is a particularly
important concept because it is a constrained resource; stronger 
measurements are in general more expensive. A particular example is the 
measurement of position by the reflection of a laser-beam~\cite{DJ,AFM}, 
a technique used in the atomic force microscope. In that case it is the 
laser power on which the measurement strength depends. Hence it is 
reasonable to consider optimization of the measurement under the assumption 
that measurement strength is fixed.

\section{Measurement and Feedback}
\label{MandF}

In classical feedback control, it is natural to consider the 
measurement process as being qualitatively different from the 
feedback process. In particular they may be distinguished by the 
fact that the measurement in each time step involves no change to 
the system Hamiltonian, and the feedback step provides no 
information. In quantum feedback, since measurement has the ability 
to affect the dynamics in ways that in classical mechanics would have 
to be attributed to a Hamiltonian, the distinction is not as 
fundamental. However, in the vast majority of quantum feedback schemes 
considered to date, it is some set of parameters describing the 
system Hamiltonian which are under the observer's control. This is 
motivated by practical considerations, since it is as of yet easiest 
experimentally to externally control
aspects of the Hamiltonian. In this case the feedback step involves
no measurement, and the observation and feedback processes may be
regarded as qualitatively different, as in the classical theory. In
view of this, the polar decomposition theorem motivates some
definitions.

By Kraus's representation theorem~\cite{Kraus}, every valid quantum
evolution (a quantum operation) may be written as a POVM given by a
set of operators $\Omega_n$, where the probability of each outcome
is $P(n)=\mbox{Tr}[\Omega_n^\dagger \Omega_n \rho]$ and the state
resulting from each outcome is $\rho_n = \Omega_n \rho
\Omega_n^\dagger/P(n)$. The only constraint on the $\Omega_n$'s is that
$\sum_{n}\Omega_n^\dagger \Omega_n =1$. However, from the polar
decomposition theorem, each of the operators $\Omega$ may be written
as the product of a unitary operator and a positive operator, so that
\begin{equation}
  \Omega_n = U_n\sqrt{\Omega_n^\dagger \Omega_n} .
\end{equation}
This provides a natural decomposition of a general quantum operation
in terms of measurement and feedback. Consider first the action of the
unitary operators. By themselves they do not describe the acquisition
of information, and in that sense they do not describe a measurement.
This can be seen from the fact that a unitary operator does not
change the von Neumann entropy of any state it acts upon, and
consequently extracts no information. However, unitary operations are
precisely the kind that can be applied by Hamiltonian feedback. Hence,
the unitary operators appearing in the polar decomposition may be
thought of as characterizing purely the feedback part of the quantum
operation. Note that we have written the polar decomposition so that
the action of the unitary operator follows after the action of the
positive operator, being a necessary condition for feedback.

Conversely, the positive operators characterize the acquisition of
information. They may always be written as a weighted sum of
projectors, and therefore thought of as providing partial information
about the states in the basis in which they are diagonal. When they
correspond to rank 1 projectors, they provide complete information, in
that the final state is pure. Since the unitary part has been factored
out to obtain the positive operators, we may regard these operators as
representing pure measurement; the change induced in the quantum state
is only that which is strictly necessary in order provide the
information obtained during the measurement. We note that this 
decomposition of measurements into unitary and positive operators has 
been considered before in the context of measurements of the first and 
second kind~\cite{Brag}.

From this it is clear that {\em every} quantum evolution can be
realized by a measurement in which the measurement operators are
positive, followed by a feedback step in which the Hamiltonian is
chosen to depend upon the measurement result. We see that the
observation of a single observable, considered in section~\ref{contm},
corresponds to the special case in which all the positive operators
forming the POVM are mutually commuting.

Under the above definitions damping processes, such as cavity decay
and Brownian motion, are not considered pure measurements; they are
viewed as equivalent to a fixed combination of measurement and
feedback. Since the object of feedback control is to limit the
deviations of a system from a desired state (or more, generally, from a
particular evolution - which means merely that the target state
changes with time), feedback control is essentially a damping process
(toward the target state).

The polar decomposition theorem therefore fits snugly with the
structure of Hamiltonian feedback, but it is nevertheless important to
realize that this is not the only feedback process that may be
considered in quantum mechanics.  First note that the product of two
positive operators need not be positive. Hence the evolution resulting
from a sequence of pure measurements as defined above will in general
be equivalent to a single pure measurement followed by some
Hamiltonian evolution (ie.  both measurement and Hamiltonian
feedback). This is an illustration of the fact that quantum
measurements involve `active' transformations of the states, as
opposed to the `passive' measurements of classical physics~\cite{DHJMT}.

Consider now the full evolution of a system under Hamiltonian feedback
control in a single infinitesimal time step $dt$, with initial state
$\rho$. Since all dynamical processes commute to first order, one can 
treat even continuous feedback control as alternating steps consisting 
of measurement and feedback. This is consistent with the general 
approach  of this paper which is to consider the two steps separately.
 The system evolves under its own `free' Hamiltonian, $H_0$,
(which in many cases will be the desired evolution), and is affected
by a source of environmental noise, which can be described by the
non-selective evolution generated by a POVM. The measurement is also
performed, and the feedback evolution applied. For a given measurement
result $n$, we may write the full evolution as
\begin{equation}
  \tilde{\rho}_n = e^{-i(H_n+H_0 )dt}P_n (\sum_j W_j \rho W_j^\dagger)
  P_n e^{i(H_n + H_0)dt}
\end{equation}
where the tilde indicates that we have not bothered to normalize the
final state, and $W_j$ are the operators describing the (undesirable)
action of the environment. Since all the operators always commute to first
order in $dt$, we have combined the free Hamiltonian with the feedback
Hamiltonian in the exponential. The task of feedback control is to
choose operators $P_n$ and $H_n$ such that the evolution is closest to
the desired evolution. Before we consider this for Hamiltonian
feedback, let us examine what can be done in the absence of the
conditional unitaries, using measurement alone, and the difference
between the two kinds of feedback.

By the definition above, using measurement alone one is restricted to
POVM's in which all the measurement operators are positive, along with
some overall unitary evolution independent of the measurement results.
Now, to evaluate the efficacy of the control procedure, we must have a
`cost function' which measures how well we have achieved the control
objective, as discussed above in the introduction. Since we have a
desired `target' state, $\sigma = |\psi_{\ms{T}}\rangle
\langle\psi_{\ms{T}}|$, in mind at some final time (to be achieved
following a single measurement, or a series of measurements), sensible
cost functions will provide a measure of how close the final state,
$\rho_{\ms{f}}$, is to the target state. A number of measures are
possible, such as the inner product
($\mbox{Tr}[\rho_{\ms{f}}\sigma]$), the fidelity
($\mbox{Tr}[\sqrt{\sigma^{1/2}\rho_{\ms{f}}\sigma^{1/2}}]$), or the
destinguishability ($(1/2)\mbox{Tr}[\rho_{\ms{f}} - \sigma]$).
Since we are interested only in target states which are pure, the 
Fidelity is simply the square root of the inner product, so that
they provide equivalent optimization problems. Throughout this paper 
we will use these as the quantities to be optimized. 


Now, the final state resulting from averaging the results of a single
pure measurement is given by 
\begin{equation}
  \rho_{\ms{f}} = \sum_n P_n \rho P_n .
\label{puremeas}
\end{equation}
Since $P_n = P_n^\dagger$, Ando's result~\cite{Ando} states that
$\rho_{\ms{f}}$ is always majorized by $\rho$, which means that the
eigenvalues of $\rho_{\ms{f}}$ are at least as evenly distributed as
the eigenvalues of $\rho$. This means that the von Neumann entropy of $\rho_{\ms{f}}$ is always at least as large as the entropy of 
$\rho$. Another way of putting this is that each eigenvalue of
$\rho_{\ms{f}}$ is some weighted average of one or more of the
eigenvalues of $\rho$. 

It follows almost immediately from the above results, that the fidelity
of the final state cannot be any larger than the maximum eigenvalue,
$\lambda_{\ms{max}(\rho)}$, of the initial state $\rho$. To see this
we first note that since all the eigenvalues of the final state,
$\lambda_j$ are a weighted average of the eigenvalues of $\rho$, none
can be larger than the largest eigenvalue of $\rho$.  Now, writing
the fidelity in terms of the eigenvectors of $\rho_{\ms{f}}$,
$|\phi_j \rangle$ we have
\begin{equation}
  \langle\psi_{\ms{T}} |\rho_{\ms{f}}|\psi_{\ms{T}}\rangle = \sum_{j} \lambda_j
  |\langle\phi_j |\psi_{\ms{T}} \rangle|^2 .
\end{equation}
Since $\sum_j |\langle\phi_j|\psi_{\ms{T}} \rangle|^2 = 1$, the fidelity is
merely a weighted average of the eigenvalues of $\rho_{\ms{f}}$, which
proves the result. In fact, choosing any basis $|\psi _i\rangle$, we
obtain the probability distribution over these states as
\begin{equation}
  \mu_i = \langle\psi_i |\rho_{\ms{f}}|\psi_i\rangle = 
          \sum_{j}\chi_{ij}\lambda_j ,
\end{equation}
where $\chi_{ij} = |\langle\phi_j |\psi_i \rangle|^2$. Since the
matrix $\chi_{ij}$ satisfies $\sum_i\chi_{ij}=1$ and
$\sum_j\chi_{ij}=1$, it is a doubly stochastic map, with the result
that the vector $\{\mu_i\}$ is majorized by the vector
$\{\lambda_j\}$, and hence the von Neumann entropy of the distribution
over any set of basis states is always at least as large as the
distribution over the eigenvectors. Another way of saying this is that
diagonal elements of a matrix resulting from a unitary transformation
performed on a diagonal matrix are always at least as uniformly
distributed as the original elements (and almost always more so).

Clearly this result for the upper bound on the final fidelity also
holds for repeated measurements, in which subsequent measurements are
{\em not} conditioned on the results of previous measurements (i.e.\
for pure measurements with no feedback). However,  it does not hold
for sequences of conditional measurements. 
In this case the initial state seen by subsequent
measurements cannot be written as the state given by averaging
over the results of previous measurements, since each final state may
have a different measurement performed on it.

It turns out that if we allow ourselves an infinite measurement
strength, then the upper bound on the final entropy derived above can
always be achieved in the limit of an infinite number of measurements.
To see this one simply follows the procedure of Aharonov and Vardi, 
referred to as the `inverse quantum Zeno effect', developed in reference~\cite{Aharonov}. Consider first an initial pure state, 
$|\psi\rangle$, and the projector, $P_\varepsilon = |\varepsilon\rangle
\langle\varepsilon|$ onto the state
\begin{equation}
  |\varepsilon\rangle = (\cos(\varepsilon)|\psi\rangle + 
                        \sin(\varepsilon)|\psi_{\ms{T}}\rangle)
\end{equation}
For $\varepsilon=0$ this is the initial state, and for
$\varepsilon=\pi/2$ this is the final state. For any value in between
this state represents a rotation through an angle $\varepsilon$ from
the initial state to the final state. If $P_\varepsilon = 
|\varepsilon\rangle \langle\varepsilon|$ makes
up one of the operators describing the measurement, the probability of
{\em failing} to obtain the state $|\varepsilon\rangle$ is
\begin{equation}
  P(\varepsilon) = \sin^2(\varepsilon) = \varepsilon^2 + \ldots
\end{equation}
If we succeed in obtaining the state $|\varepsilon\rangle$, then in
that measurement step we have succeeded in rotating the state through
$\varepsilon$ toward the desired state. We can attempt to rotate the
state through the full $\pi/2$ radians by choosing $\varepsilon =
\pi/(2M)$ and using $M$ measurements. Since the probability of failure
at each step is then second order in $(1/M)$, while the number of
steps scales only as $M$, as the number of steps tends to infinity,
the total probability of failure tends to zero, and in this limit one
achieves the desired rotation. To see that this achieves the upper
bound when the initial state is mixed, we choose the projector so as
to rotate the eigenvector of $\rho$ corresponding to the maximum
eigenvalue to the desired state. For each measurement the corresponding
POVM is then given, in general, by $P_{\varepsilon_i} + \sum_l \Omega_l^\dagger \Omega_l = 1$, where $P_{\varepsilon_i}$ is the projector for the $i^th$ measurement, and the $\Omega_l$ are arbitrary. The final state may then be 
written
\begin{eqnarray}
  && \rho_{\ms{f}} = \prod_i^M {\cal L}_{\varepsilon_i}\left[ \lambda_1|1\rangle\langle 1| \right] + \prod_i^M {\cal L}_{\varepsilon_i}\left[\sum_{n=2}^{N}\lambda_n |n\rangle\langle n| \label{rhofm}\right] \\
  && {\cal L}_{\varepsilon_i}\left[A\right] = P_{\varepsilon_i} A P_{\varepsilon_i} + \sum_{l} \Omega_l A \Omega^\dagger_l
\end{eqnarray}
where $\rho |n\rangle = \lambda_n |n\rangle$, with $\lambda_1$ the maximum eigenvalue, and $A$ is an arbitrary operator. As the number of measurements tends to infinity, the first term on the RHS of Eq.(\ref{rhofm}) becomes
\begin{equation}
  \lim_{M\rightarrow\infty}{\cal L}_{\varepsilon_i} \lambda_1|1\rangle\langle 1| = \lambda_1 |\psi_{\ms{T}}\rangle \langle \psi_{\ms{T}}|  .
\end{equation}
Since this term contributes $\lambda_{\ms{max}}(\rho)$ to the fidelity, and since the other pure states making up the final density matrix cannot contribute negatively, the upper bound is achieved.

What happens when we allow ourselves infinite measurement strength,
and sequences of measurements in which subsequent measurements are
{\em conditioned} on previous results (i.e.\ measurement-only
feedback)? In that case it is clear that the desired state can always
be obtained with certainty; one begins by making a projection
measurement in an arbitrary basis, which results in a set of pure
states. Then the above procedure is used to rotate the resulting state
to the target. 

When infinite strength is available, measurement-only feedback is
equivalent to Hamiltonian feedback, since both allow any state to be
created. However, in many continuous feedback control
applications, the strength of the measurement is unlikely to be so
much stronger than either the environmental noise or the free
system dynamics that it can be used in this fashion in place of
Hamiltonian feedback~\cite{CQED,CHP,Tittonen98}. With measurements 
of finite strength the outcomes are necessarily random, so that 
Hamiltonian feedback cannot be simulated reliably. One can expect 
therefore that real applications will find the use of Hamiltonian 
feedback invaluable.



\section{Measurement: Maximal Information and Minimal Disturbance}
\label{maxmin}
In Section~\ref{sec:ms} we introduced the concept of the information
provided about the system by a measurement, and this involved
specifically the information regarding the state {\em resulting} from
the measurement, a definition motivated by feedback control. This in
turn motivated the definition of the strength of a measurement
(e.g. $s_{\ms{v}}$ or $s_{\ms{p}}$), important because it constitutes a
natural constraint when considering the optimization of control
strategies. However, the actual information provided by a given
measurement is not only a function of the measurement strength, but
also the state of the system immediately prior to the measurement. As a
result, once the available measurement strength is known, one can ask
how to optimize the information provided by the measurement given the
current state of the system. This defines the concept of a measurement
returning maximal information (for a fixed measurement strength).

In addition to providing information, quantum measurements can also
introduce noise, a statement which we will now make precise. Consider
first a classical system driven by noise. One can characterize the
extent of the noise in some time interval by the increase in the
entropy of the phase space probability distribution for the system
state which is given by averaging over the noise realizations. This tells
us how much we expect the noise to spread out the system in phase
space in that time interval, and characterizes our uncertainty about
the future state of the system resulting from the noise. Now consider
a classical measurement. Since the initial state is uncertain (or else
we would not need to make the measurement), the result of the
measurement is random, and as a result one's state of knowledge changes
in a random fashion. However, this random change should not be
considered noise, since if one averages over all the possible
measurement results (all the possible random changes) the probability
distribution for the state of system remains unchanged. This is the
sense in which classical measurement introduces no noise into the
system.

Now consider a quantum system driven by noise. The equivalent of the
phase space distribution is the density matrix. In the same manner one
can characterize noise by the resulting increase in the von Neumann
entropy of the density matrix resulting from averaging over the
possible noise realizations. One can therefore characterize the noise
introduced by a quantum measurement by calculating the increase in the
von Neumann entropy (or alternatively the decrease in the purity) 
of the density matrix which results from averaging
over the possible measurement results. While we saw above that in the
classical case the measurement introduces no noise, this is not, in
general true for quantum measurement. In terms of the von Neumann 
entropy, the excess noise introduced by a measurement is
\begin{equation}
  N_{\ms{e}}^{\ms{v}} = S_{\ms{v}}(\rho_{\ms{f}}) - S_{\ms{v}}(\rho) .
\end{equation}
Defining it in terms of the purity we have
\begin{equation}
  N_{\ms{e}}^{\ms{p}} = \mbox{Tr}[\rho^2] - \mbox{Tr}[\rho_{\ms{f}}^2] .
\end{equation}
This makes precise the intended
meaning of our initial statement that quantum measurements can introduce
noise. Note that this has nothing to do with the Heisenberg
uncertainty principle --- what concerns us here is the uncertainty of
the future quantum state, and not the uncertainty of some set of
observables for a given state. Recall that this is because the object
of the control is the state of the system, and it is up to the
observer to decide what the desired state is. Whether it be a minimal
uncertainty state in the sense of the Heisenberg uncertainty principle
is immaterial.


Let us consider first the question of minimizing the disturbance due
to the measurement. Recall that for a pure measurement, the 
evolution given by averaging over the measurement results is given by
Eq.~(\ref{puremeas}), where all the $P_n$ are positive operators. Once
again invoking Ando's result, we have that the von Neumann entropy of
the final state is never decreased by the measurement. Measurements
which minimize noise are therefore the measurements which leave the
von Neumann entropy unchanged. These measurements are in this sense
most like classical measurements. A set of measurements
satisfying this criterion are those in which all the $P_n$ commute
with the initial density matrix. In this case we have
\begin{equation}
  \rho_{\ms{f}} = \sum_n P_n \rho P_n  = \sum_n P_n^2 \rho = \rho .
\end{equation}
In the language of continuous measurements, since the operators $P_n$
are diagonal in the eigenbasis of the observable, this means choosing
to measure an observable which shares an eigenbasis with the density
matrix.

On a practical note, for continuous observation, measuring in
the eigenbasis of the density matrix involves continuously changing 
the measured observable (note that such a process has been considered 
previously in the context of adaptive measurements~\cite{adapt}). In 
many situations this 
flexibility may be only partially available, or not at all. However,
the above analysis indicates that for the purposes of noise
minimization, one should choose the measured observable to be that in
which the system is diagonal, or nearly diagonal, for the longest time
during the period of control. In fact, this introduces the possibility
that in certain cases it may be desirable to turn off measurement for
periods in which the system occupies states which have large
off-diagonal elements in the eigenbasis of the observable. Of course,
the resulting noise reduction would have to be balanced against the
accompanying loss of information.

Maximizing the information for a fixed measurement strength is a much
more difficult problem. Here we will examine a specific example for the
continuous measurement of a two-state system. In the formulation of
continuous measurements that was discussed in Section~\ref{contm}, we
used measurement operators where each was a sum over an infinite
number of projectors. For a two state system it is possible to obtain
the same result (i.e. the same continuous measurement driven by
Gaussian noise) by using a formulation with only two measurement
operators where each is the sum over only two projectors. To obtain
a continuous measurement of a given observable the POVM is given by
$\Omega_0^2 + \Omega_1^2 = 1$ where the measurement operators are
\begin{eqnarray}
  \Omega_0 & = & \sqrt{\kappa}|0\rangle \langle 0| +
                 \sqrt{1-\kappa}|1\rangle \langle 1| , \\
  \Omega_1 & = & \sqrt{\kappa}|1\rangle \langle 1| + 
                 \sqrt{1-\kappa}|0\rangle \langle 0|
\end{eqnarray}
in which $\kappa = 1/2 + \sqrt{kdt}$, and $|0\rangle$ and $|1\rangle$
are the eigenstates of the observable.  In each time step $dt$ this
produces one of two results. The sum of these, in a time interval
$\Delta t = Ndt$ in which $N$ results are obtained, is naturally
governed by the binomial distribution. In the limit of large $N$ (and
infinitesimal $\Delta t$) this tends to a Gaussian, and one obtains
the measurement record (Eq.(\ref{mrec})) and SME (Eq.(\ref{SME}))
given in Section~\ref{contm}, where the measured observable  
$Q = |0\rangle \langle 0| - |1\rangle \langle 1|$. We can alternatively 
think of this measurement as a single-shot measurement, and in that case 
$\kappa$ can take any value between zero and one. Note that when 
$\kappa=0$ or $\kappa=1$ the measurement is one of infinite strength. As 
$\kappa$ becomes closer to $1/2$, the strength reduces, and for 
$\kappa=1/2$ the measurement provides no information.

We can obtain measurements of all possible observables by applying to 
the measurement operators an arbitrary rotation over the Bloch sphere, 
given by the unitary transformation
 \begin{eqnarray}
   U(\theta,\phi)|0\rangle & = & \cos(\theta/2) |0\rangle 
                                + e^{i\phi}\sin(\theta/2) |1\rangle \\
  U(\theta,\phi)|1\rangle & = & \cos(\theta/2) |1\rangle 
                                - e^{-i\phi}\sin(\theta/2) |0\rangle.
\end{eqnarray}
Recall that this unitary transformation of the measurement operators
preserves the measurement strength as defined in section~\ref{sec:ms}. 
Without loss of generality, we can choose the initial density matrix 
to be diagonal, and write it as $\rho = p |0\rangle \langle 0|+ (1-p)|1\rangle \langle 1|$. One can then obtain an analytic expression for the final average purity, which is given by
\begin{equation}
    I_{\ms{f}}^{\ms{p}} \equiv 
    1 - u_{\ms{p}} = \sum_{n=0}^1
    \frac{\mbox{Tr}[(U\Omega_n^\dagger U^\dagger\rho U\Omega_n
    U^\dagger)^2]}{\mbox{Tr}[U\Omega_n^\dagger U^\dagger\rho U\Omega_n
    U^\dagger]} .
\end{equation}
This expression is fairly complex, and we will not need it here. (For a  
detailed analysis of this expression, including analytic 
expressions for general two-outcome measurements on two-state systems, 
the reader is referred to~\cite{FJ}). It is explicitly independent of
$\phi$, as one would expect, since it is $\theta$ alone which gives
the angle (on the Bloch sphere) between the basis in which the density
matrix is diagonal, and the basis of the measured observable. The
final average purity is then explicitly dependent on the three
parameters, $p$, $k$ and $\theta$, and we are concerned with
maximizing this with respect to $\theta$. When the measurement is 
non-trivial ($\kappa \not= 0.5$), the strength of the 
measurement is finite ($\kappa \not= 0,1$) and the initial state is 
impure ($p \in (0,1)$), one finds that the location
of the maximum is independent of $p$ and $k$, and occurs for $\theta =
\pi/2$. This means that on average the maximum information is obtained
about the final state (for fixed measurement strength) when the basis
of the measured observable is maximally {\em different} from the
eigenbasis of the density matrix (ie. if the density matrix is a
mixture of $\sigma_z$ eigenstates, then one should measure $\sigma_x$
or $\sigma_y$). 

\begin{figure}[h]
\centerline{\psfig{file=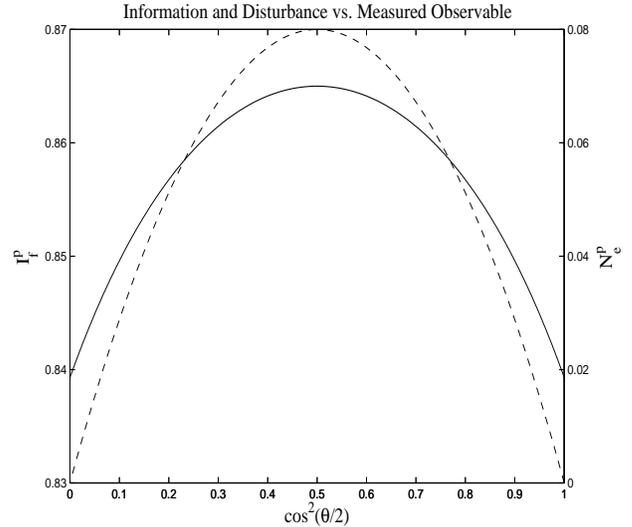,width=3.25in,height=2.8in}}
\caption{\narrowtext Here we plot the information obtained about 
the final state (solid line), characterized by the final average 
purity, $I_{\ms{f}}^{\ms{p}}$, and the excess noise introduced by the measurements (dashed line), $N_{\ms{e}}^{\ms{p}}$, against the 
measured observable, parametrized by the angle $\theta$. 
The parameters are $p=0.1$ and $\kappa=0.75$}
\label{fig1}
\end{figure}

We see then, that at least for a two-state system, the minimal
disturbance is obtained when the measured observable has the same
eigenbasis as the density matrix, and the maximal information is
obtained when the two bases are maximally different. Thus we obtain 
the result that, at least for a two-state system, there is a trade-off 
between information and disturbance in quantum feedback control (in 
contrast to classical feedback control). This trade-off for finite 
strength quantum measurements is also of interest from a purely 
fundamental point of view, and this is explored in detail in 
reference~\cite{FJ}. We plot both the excess noise introduced by the 
measurement, and the average final purity resulting from the measurement 
as a function of $\theta$ in Fig.~\ref{fig1}. For a fixed measurement 
strength one therefore has the choice between choosing a measurement to 
minimize the noise, and consequently obtain better control of the system 
(in that the system will fluctuate less around the desired value), or 
obtain a more accurate knowledge of the system at the expense of increased 
noise. Which is most desirable may well depend upon the current state of 
knowledge. For example, if the state of the system is poorly known, 
perhaps early on in the control process, then it may prove desirable to 
obtain information more quickly, at the expense of introducing extra 
noise, since the large uncertainty will be the major factor in reducing 
the effectiveness of the feedback. However, once the observer's knowledge 
is sufficiently sharp, it may prove more effective to reduce the noise at 
the expense of some added uncertainty. In Section~\ref{sim} we will 
present simulations to show how this information trade-off affects the 
performance of feedback control in a two-state system.

\section{Optimal Hamiltonian Feedback}
\label{opham}
Optimizing the Hamiltonian feedback for each feedback step consists in
finding the feedback Hamiltonian which maximizes the fidelity with the
target state, given the initial state of the system, which is fairly straightforward. Before we consider optimizing an infinitesimal 
Hamiltonian, which is necessary for continuous feedback, let us find 
the optimal unitary transformation. This result would be useful if a 
finite time were available for the feedback step, and feedback strength 
was not an important constraint, so that a `rotation' through any angle 
could be performed by the feedback Hamiltonian.

Denoting the state of the system at the beginning of the feedback step
as $\rho$, the fidelity with respect to the target state at the end of
the feedback step is given by
\begin{equation}
  F(\rho_{\mbox{f}},\sigma) =  \mbox{Tr}[\sqrt{\rho^{1/2}\sigma\rho^{1/2}}] ,
\end{equation}
where $\sigma = |\psi_{\ms{T}}\rangle\langle\psi_{\ms{T}}|$ is the
target state, and $U$ is the unitary transformation constituting the
feedback. We wish to find $U$ to maximize
$F(\rho_{\ms{f}},\sigma)$. Observing first that, for arbitrary $A$
and unitary $V$, 
\begin{eqnarray}
  \max_{V}|\mbox{Tr}[AV]| 
             & = & \max_{V}|\mbox{Tr}[\sqrt{A^\dagger A} V' V]| \nn \\
             & = & \max_{V}|\sum_{j}\sigma_j(A)e^{i\theta_j}| \nn \\
             & = & \mbox{Tr}[\sqrt{A^\dagger A}]
\end{eqnarray}
where we have used the polar decomposition theorem for $A$ ($A =
\sqrt{A^\dagger A}V'$), and the $\sigma_j(A)$ are the eigenvalues of
$\sqrt{A^\dagger A}$. Setting $A = \rho_{\ms{f}}^{1/2}\sigma^{1/2}$,
this gives
\begin{eqnarray}
  F(\rho_{\ms{f}},\sigma) 
  & = & \max_{V}|\mbox{Tr}[(U\rho U^\dagger)^{1/2}\sigma^{1/2}V]| \nn \\
  & = & \max_{V}|\mbox{Tr}[U\rho^{1/2}U^\dagger\sigma^{1/2}V]| \nn \\
  & \leq & \sum_{j}\lambda_j(\rho)^{1/2}\lambda_j(\sigma)^{1/2}
\end{eqnarray}
where the final inequality uses the result by von Neumann~\cite{vonN}. In
the last line $\lambda_j(\rho)$ and $\lambda_j(\sigma)$ are the
eigenvalues of $\rho$ and $\sigma$ respectively, ordered such that the
largest eigenvalue of $\rho$ multiplies the largest eigenvalue of
$\sigma$, the second largest the second largest, and so on down to the
smallest eigenvalue of both states. Now we need merely realize that we
can achieve the upper bound by choosing $U$ so as to diagonalize
$\rho$ in the basis of $\sigma$, re-ordering the basis states such
that the largest eigenvalue of $\rho$ is attached to the eigenstate of
$\sigma$ with the largest eigenvalue. Writing the eigenstates of $\sigma$
as $|\sigma_j\rangle$ (with eigenvalues ordered by size), and those of
$\rho$ as $|\rho_j\rangle$ (similarly ordered), then the explicit
construction for the optimal $U$ is 
\begin{equation}
  U =  \sum_j |\sigma_j\rangle \langle\rho_j|.
\end{equation}
For continuous observation, each feedback step acts only for an
infinitesimal time, $dt$. In this case the final state,
$\rho_{\ms{f}}$, is given by
\begin{equation}
  \rho_{\ms{f}} = \rho - i[H,\rho]\Delta t - \half [H,[H,\rho]] (\Delta t)^2 + \ldots ,
\end{equation}
where $H$ is the feedback Hamiltonian. The fidelity of the final state with the target state is therefore
\begin{eqnarray}
 \langle \psi_{\ms{T}}| \rho_{\ms{f}}|\psi_{\ms{T}} \rangle & = & \langle
 \psi_{\ms{T}}|\rho|\psi_{\ms{T}} \rangle \nn \\
  & - & i\langle \psi_{\ms{T}}|[H,\rho]|\psi_{\ms{T}} \rangle \Delta t \nn \\
  & - &  \half \langle \psi_{\ms{T}}|[H,[H,\rho]]|\psi_{\ms{T}}
 \rangle (\Delta t)^2 + \ldots
\label{fidterms}
\end{eqnarray}

The first term is fixed, so to optimize the fidelity in the
infinitesimal time step we should maximize the coefficient of $\Delta t$,
being the dominant term. If the target state commutes with $\rho$, then 
this term vanishes for all $H$, so that we cannot choose a Hamiltonian that 
will cause an increase in the fidelity that is first order in time. If this 
situation occurs only for vanishingly small times, then it will make 
effectively no difference to the feedback performance. However, in those 
special situations in which the intrinsic dynamics preserves the commutivity 
of $\rho(t)$ and the target state, it can be important to choose a 
Hamiltonian which maximizes the term which is second order in time. One should note, however, that if one has freedom to choose the measurement basis, one can always choose a basis which disturbs the state so as to break the commutivity of $\rho(t)$ with the target state, eliminating the need to consider the second order term in the time evolution.

The maximization must be performed under a reasonable constraint on
the eigenvalues of $H$ (i.e. a constraint that captures the concept of
a limitation on the strength of feedback). A number of suitable
constraints are possible, such as a restriction on the maximum
eigenvalue of $H$, the sum of the norms of the eigenvalues, the sum of
the squares of the eigenvalues, etc. Here we choose to use the
last of these constraints, namely
\begin{equation}
  \sum_n \lambda_n(H)^2 \leq \mu .
\end{equation}
To maximize the coefficient of $\Delta t$ in Eq.(\ref{fidterms}), we first 
note that
it may be written as the operator inner product $\mbox{Tr}[H A]$, where 
\begin{eqnarray}
  A  & = & i |\psi_{\ms{T}} \rangle \langle v | 
         - i |v \rangle  \langle \psi_{\ms{T}}|, \nn \\
  |v \rangle & = & \rho |\psi_{\ms{T}} \rangle
\end{eqnarray}
The maximum of the inner product, under the condition that the norms
of the operators are constrained, occurs when the operators are
aligned: $H = c A$, where $c$ is in general a complex number, but real
in this case to preserve the Hermiticity of $H$. With this inner
product the norm of $H$ is
\begin{equation}
  \mbox{Tr}[H^2] = \sum \lambda_n(H)^2 .
\end{equation}
Naturally, we take the maximum value allowed under the constraint, setting
$\mbox{Tr}[H^2] = \mu$. This fixes the magnitude of the proportionality
constant $c$, and results in the following explicit construction for
the Hamiltonian 
\begin{equation}
  H = i \chi [ |\psi_{\ms{T}} \rangle \langle \psi_{\ms{T}}|, \rho ].
\end{equation}
where
\begin{equation}
    \chi = \sqrt{\frac{\mu}{a - b^2}} ,
\end{equation}
with
\begin{eqnarray}
  a & = & \langle \psi_{\ms{T}}| \rho^2 |\psi_{\ms{T}} \rangle \\
  b & = & \langle \psi_{\ms{T}}| \rho |\psi_{\ms{T}} \rangle
\end{eqnarray}

It now remains to maximize the coefficient of $(dt)^2$ in
Eq.(\ref{fidterms}). Recall this is only required under the condition
that the first term is zero, which implies that $\rho|\psi_{\ms{T}}\rangle = \lambda_{\ms{T}}|\psi_{\ms{T}}\rangle$. In this case the expression for
the coefficient may be written as
\begin{equation}
   \langle \psi_{\ms{T}}|H(\rho - \lambda_{\ms{T}}I)H|\psi_{\ms{T}} \rangle .
\end{equation}
Now, denoting the eigenvalues of $\rho$ by $\lambda_n$ (ordered in decreasing order), the eigenstates by $|n\rangle$, and denoting the target 
state as the eigenvector with $n=M$, the above expression becomes
\begin{equation}
  \langle \psi_{\ms{T}}|H\rho H|\psi_{\ms{T}} \rangle = \sum_{n=1}^{N} (\lambda_n - \lambda_{M})|\langle n | H | M \rangle|^2 .
\end{equation}
Now, the constraint on the Hamiltonian may be written
\begin{eqnarray}
  \mbox{Tr}[H^2] & = & \mbox{Tr}[\sum_{n}|n\rangle\langle n| H
  \sum_{m}|m\rangle\langle m| H] \nn \\
    & = & \sum_{nm} |\langle n | H | m \rangle|^2 = \mu ,
\end{eqnarray}
being a constraint on the sum of the square magnitudes of the elements
of $H$. Only the subset $|\langle n | H | M \rangle|^2$ of these, with 
$n\neq M$, 
contribute to the expression to be maximized. To obtain the 
maximum value for the expression, we must therefore set all the 
elements of $H$ which do not contribute to it to zero, this allowing 
the contributing elements to be as large as possible. The constraint then becomes
\begin{equation}
  \sum_{n\neq M}^{N} |\langle n | H | M \rangle|^2 = \mu/2 ,
\end{equation}
where the factor of one half is enforced by the Hermiticity of $H$.
The expression can now be seen as an average of the eigenvalues of
$\rho$ over the `distribution' $P(n) = |\langle n | H | M \rangle|^2$,
which is normalized to $\mu$ by the constraint. The maximum value is
therefore achieved when all the weight of the distribution is placed
on the term with the largest eigenvalue. The solution is therefore
\begin{equation}
  |\langle 1 | H | M \rangle|^2 = |\langle M | H | 1 \rangle|^2 = \mu/2,
\end{equation}
with all other elements zero. The explicit construction for the 
optimal $H$ being
\begin{equation}
  H = \sqrt{\mu/2} (|1\rangle \langle \psi_{\ms{T}}| +
  |\psi_{\ms{T}}\rangle \langle 1|) .
\end{equation}
Note that this assumes that the target state is orthogonal to $|1\rangle $, being the eigenvector with the largest eigenvalue. If the $|1\rangle$ is the target state, then there exists no Hamiltonian evolution which will increase the fidelity, since the fidelity is the maximum it can be given the current purity of $\rho$. In that case we are free to set $H = 0$ for that time step.

It is worth noting that since the magnitude of the feedback Hamiltonian and the strength of the continuous observation are uniformly bounded, the evolution of the system is continuous. Given this, since the Feedback Hamiltonian is a continuous function of the system state, it is intuitively clear that the feedback algorithm is well-defined (and continuous) in the continuum limit for almost all sample paths.

We now have a feedback algorithm which can be used in conjunction 
with a measurement strategy for feedback control. In the next section we will implement such a strategy for the control of a two-state system.

\section{Feedback Control of a Two-State System}\label{sim}
In the previous sections we have considered the measurement and Hamiltonian feedback parts of the control problem separately. This resulted in a straightforward choice for a Hamiltonian feedback algorithm, but did not result in a clear choice for the measurement strategy. This was because we were able to identify in the measurement process a trade-off between information and disturbance. Because of this, the optimal measurement strategy for a given application is likely to depend upon the relative strengths available for the measurement and feedback. For example, if the feedback Hamiltonian is relatively strong, then it is likely that it will be able to effectively counter the disturbance introduced by the measurement, and therefore the measured observable should be chosen to provide maximal information and the expense of maximal disturbance. When this is not the case, the most desirable measurement is likely to be that which introduces less disturbance at the expense of providing reduced information.

To examine the performance of a feedback control algorithm we must run the algorithm many times in order to obtain the average behavior. This is computationally very expensive, and so we use massively parallel super-computers which are ideal for this task. The results we present here are obtained by averaging one thousand realizations of the control algorithm. 

To provide a simple example of feedback control we consider a spin-half system precessing in a magnetic field aligned along the z-axis. In the absence of any noise, a spin aligned originally along the x-axis would rotate at a constant angular velocity around the z-axis, and we take this to be the desired (target) behavior. To provide the control problem, we subject the spin to noise which dephases it around the z-axis (this could arise from fluctuations in the magnetic field). The master equation describing the free (but noisy) evolution of the spin is thus given by
\begin{equation}
  \dot{\rho} = -i\hbar \omega [\sigma_z,\rho] - \beta [\sigma_z,[\sigma_z,\rho]] ,
\end{equation}
where $\omega$ is the precession frequency in the magnetic field and $\beta$ is the strength of the dephasing noise. To implement feedback control we allow the observer (who is also naturally the controller) to measure the spin along an arbitrary spin direction ${\bf v}(t)$, with measurement constant $k$, and apply a feedback Hamiltonian, $H_{\mbox{\scriptsize fb}}(t)$, obtained using the algorithm presented in the previous section. The full evolution of the controllers' state of knowledge, including the measurement and feedback, is therefore
\begin{eqnarray}
  d\rho & = & -i\hbar [\sigma_z + H_{\mbox{\scriptsize fb}}(t),\rho] dt
              - \beta [\sigma_z,[\sigma_z,\rho]] dt \nn \\
           && - k [\sigma_{{\bf v}(t)},[\sigma_{{\bf v}(t)},\rho]] dt \nn \\
	   && + \sqrt{2k} (\sigma_{{\bf v}(t)}\rho + \rho\sigma_{{\bf v}(t)} -  2\mbox{Tr} [\sigma_{{\bf v}(t)}\rho]\rho) dW
\end{eqnarray}

\begin{figure}[h]
\centerline{\psfig{file=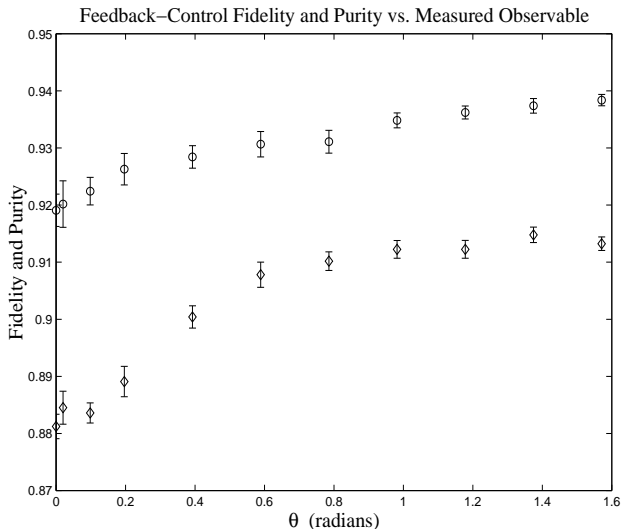,width=3.25in,height=2.8in}}
\caption{\narrowtext The average purity (diamonds) and fidelity (circles) of a feedback control algorithm for different measurement angles $\theta$. The parameters are precession frequency $\hbar\omega = \pi$, measurement constant $k=2$, noise strength $\beta = 0.4$ and feedback strength $\mu = 10$.}
\label{fig2}
\end{figure}

We now simulate the dynamics resulting from the feedback control loop for different values of $\theta$, being the angle between the eigenbasis of the instantaneous system density matrix and the instantaneous measured observable, as discussed in Section~\ref{maxmin}. For these simulations the strength of the magnetic field is such that $\hbar\omega=\pi$, so that the spin rotates once in a time interval $t=1$. The noise strength is $\beta = 0.4$, and the parameters for the control loop are $k=2$ and feedback strength $\mu=10$. We start the system in a pure state with the spin pointing along the x-direction, and evolve the controlled dynamics for a duration of $t=2$ (the purity and fidelity settle down to their steady state behavior by approximately $t=0.8$). Averaging the fidelity and purity over the full length of the run, for different values of $\theta$ we obtain figure~\ref{fig2}. Examining the dependence of the purity on $\theta$, we find what we expect from the discussion in Section~\ref{maxmin}. That is, the average purity of the system increases with $\theta$, achieving a maximum at $\theta=\pi/2$. This reflects the fact that, on average, measurements with a larger value of $\theta$ extract information from the system at a faster rate. 

The behavior of the fidelity, in this case, is similar to that of the purity. As $\theta$ increases, the feedback is sufficient to ensure that even though we can expect the noise to increase with $\theta$, the increase in purity has more of an effect on the fidelity than the noise. The result is that, with these resources, it is best to choose $\theta = \pi/2$ (so as to measure in a basis maximally different from that which diagonalises the density matrix). However, from our previous analysis of the trade-off between information and disturbance, we cannot always expect this to be the case.

\section{Conclusion}
\label{conc}
In this paper we have considered the problem of controlling a quantum
system in real time using feedback conditioned on information obtained
by continuous observation. The question of how to effect the best
control given the system dynamics (including environmental noise), and
constraints on available resources is highly non-trivial. Here we
considered a simplified problem in which we examine the measurement
process and the resulting Hamiltonian feedback separately. Our purpose
was both to examine what concepts are motivated by the feedback
control problem, and to explore the question of optimization in
this simplified problem. A concept which arises immediately in
considering feedback control is the strength of a measurement. This
strength quantifies the amount of information which a measurement
provides. Previous definitions of the information provided by quantum
measurements have focussed on information regarding the state prior to
the measurement. Here we have argued that it is the information
regarding the state {\em resulting} from the measurement which is
relevant to quantum feedback control, and introduced a concept 
of measurement strength accordingly.

Since measurements disturb quantum systems, it is important to
understand how this relates to feedback control. We showed how it is
possible to quantify the concept of the noise introduced by
measurements in a way that is relevant to feedback control. One finds
that while classical measurements do not introduce noise, quantum
measurements do, in general, although it is possible, at least in
principle, to make continuous quantum measurements that are noise
free.

Having arrived at precise concepts of information and disturbance, we
examined the special case of continuous measurements performed on a
two-state system, and found that maximization of information, and
minimization of noise were mutually exclusive goals, implying the
existence of an information-disturbance trade-off in quantum feedback
control. This highlights the complexity of the control problem.

We also considered the Hamiltonian feedback part of the control
process. Defining the cost function as the fidelity with a target
state, and the feedback strength as the norm of the Hamiltonian, we
were able to obtain the Hamiltonian generating the optimal
instantaneous feedback. 

Here we explicitly consider control realized by choosing dynamics 
conditional upon a measurement process. Because of this one can refer 
to this technique as using a classical controller, since it works by 
taking a classical process (the measurement record) and altering the 
system Hamiltonian accordingly, all of which can be achieved by a 
classical system. It is therefore worth noting that, so long as we 
are considering the dynamics of the controlled system alone to be 
the important quantity, this is equivalent to control which is realized 
by connecting the system, via an interaction Hamiltonian, to another 
quantum system, where this second system is large enough to be treated 
as a bath~\cite{HMWPhD}. In general, using a second quantum system in 
this fashion may be referred to as using a quantum controller. When 
the quantum controller is finite dimensional and restricted in its 
dynamical response time, one can expect the performance of classical 
and quantum controllers to be somewhat different, and this is an 
interesting area for future work.

The question of how to best design feedback strategies to control noisy
quantum systems is a complex one. However, the study of this problem
will help us to understand better how quantum measurement may be
exploited in the manipulation of quantum systems, and as quantum
technology advances we can expect that this question will become
increasingly important in practical applications.

\section*{Acknowledgments}
The authors would like to thank Howard Barnum, Tanmoy Bhattacharya, Chris Fuchs and Salman Habib for helpful discussions. KJ would like to thank Professor Sze Tan for hospitality during a visit to the University of Auckland where part of this work was carried out. This research was performed in part using the resources located at the Advanced Computing Laboratory of Los Alamos National Laboratory.

\end{multicols}
\widetext
\begin{multicols}{2}

\end{multicols}

\begin{thebibliography}{10}
\vspace{-1.5cm}

\bibitem{CQED} H. Mabuchi, J. Ye, and H.J. Kimble, Appl. Phys. B, 1095
(1999), Eprint: quant-ph/9805076.

\bibitem{ion} B.E. King, C.S. Wood, C.J. Myatt, Q.A. Turchette,
D. Leibfried, W.M. Itano, C. Monroe, and D.J. Wineland,
Phys. Rev. Lett. {\bf 81}, 1525 (1998), Eprint: quant-ph/9803023.

\bibitem{Bel1} V.P. Belavkin, {\em Automat. Remote Control}, {\bf
44} 178 (1983).

\bibitem{qfb1} H.M. Wiseman and G.J. Milburn, Phys. Rev. Lett. {\bf
70}, 548 (1993); Phys. Rev. A {\bf 49}, 2133 (1994); {\em ibid} {\bf
49}, 5159(E) (1994); {\em ibid} {\bf 50}, 4428(E) (1994).

\bibitem{sloss} J.J. Slosser and G.J. Milburn,
Phys. Rev. Lett. {\bf 75}, 418 (1995); P.~Tombesi and D.~Vitali,
Appl. Phys. B {\bf 60}, S69 (1995); Phys. Rev. A {\bf 51}, 4913
(1995); P. Goetsch, P. Tombesi, and D. Vitali, Phys. Rev. A {\bf 54},
4519 (1996); D.B. Horoshko and S.Ya. Kilin, Phys. Rev. Lett. {\bf 78},
840 (1997).

\bibitem{dunn} J.A. Dunningham, H.M. Wiseman, and
D.F. Walls, Phys. Rev. A {\bf 55}, 1398 (1997); S. Mancini and
P. Tombesi, Phys. Rev. A {\bf 56}, 2466 (1997); S. Mancini, D. Vitali,
and P. Tombesi, Phys. Rev. Lett. {\bf 80}, 688 (1998), 
Eprint: quant-ph/9802034.

\bibitem{hof} H.F. Hofman, G. Mahler, and O. Hess,
Phys. Rev. A {\bf 57}, 4877 (1998).

\bibitem{taub} M.S. Taubman, H.M. Wiseman, D.E. McClelland,
and H.-A. Bachor, J. Opt. Soc. Am. B {\bf 12}, 1792 (1995);
H.M. Wiseman, Phys. Rev. Lett. {\bf 81}, 3840 (1998), 
Eprint: quant-ph/9805077.

\bibitem{BelLQG} V.P. Belavkin, {\em Non-demilition measurement
and control in quantum dynamical systems}, In: Information
complexity and control in quantum physics, ed. A. Blaquiere, S. Diner
and G. Lochak, (Springer-Verlag, New York, 1987).

\bibitem{DJ} A. Doherty and K. Jacobs, Phys. Rev. A {\bf 60}, 2700
(1999), Eprint: quant-ph/9812004.

\bibitem{DHJMT} A.C. Doherty, S. Habib, K. Jacobs, H. Mabuchi and
  S.M. Tan, in submission, Eprint: quant-ph/9912108.

\bibitem{HMWPhD} H.M. Wiseman, Ph.D. Thesis, University of Queensland
  (1994).

\bibitem{Jacobs} O.L.R. Jacobs, {\em Introduction to Control Theory},
(Oxford University Press, Oxford, 1993).

\bibitem{Maybeck} P.S. Maybeck, {\em Stochastic Models, Estimation
and Control}, volumes II and III, (Academic Press, New York, 1982).

\bibitem{Whittle} P. Whittle, {\em Optimal Control}, (John Wiley \& Sons,
Chichester, 1996).

\bibitem{Doyle} K. Zhou with J.C. Doyle and K. Glover, {\em Robust and
    Optimal Control}, (Prentice Hall, New Jersey, 1995).
    
\bibitem{Traj} H.J. Carmichael, {\em An Open Systems Approach to Quantum Optics}, Lecture Notes in Physics m18, (Springer-Verlag, Berlin, 1993); C.W. Gardiner and P. Zoller, {\em Quantum Noise} (Springer-Verlag, Berlin, 1999).

\bibitem{MC} C. Caves and G.J. Milburn, Phys. Rev. A. {\bf 36}, 5543
  (1987).

\bibitem{KJPhD} K. Jacobs, Ph.D. Thesis, Imperial College, London
  (1998), Eprint: quant-ph/9810015.
  
\bibitem{Massar} S. Massar and S. Popescu, Phys. Rev. Lett. {\bf 74}, 1259 (1995).

\bibitem{Derka} R. Derka, V. Buzek and A.K. Ekert, Phys.Rev.Lett. {\bf 80}, 1571 (1998).
  
\bibitem{Gardiner1} C. W. Gardiner, {\em A Handbook of Stochastic Methods}, 2nd ed. (Springer-Verlag, Berlin, 1985).

\bibitem{AFM} G.J. Milburn, K. Jacobs and D.F. Walls, Phys. Rev. A {\bf 50}, 5256 (1994); K. Jacobs, I. Tittonen, H.M. Wiseman and S. Schiller, Phys. 
Rev. A {\bf 60}, 538 (1999).

\bibitem{Kraus} K. Krauss, {\em States, Effects and Operations: Fundamental Notions of Quantum Theory} (Springer, Berlin, 1983).

\bibitem{Brag} V.B. Braginsky and F.Y. Khalili, {\em Quantum Measurement} (Cambridge University Press, Cambridge, 1992).

\bibitem{Ando} T. Ando, Linear Algebra and its Applications {\bf 118}, 163 (1989).

\bibitem{Aharonov} Y. Aharonov and M. Vardi, Phys. Rev. D {\bf 21}, 2235 (1980).

\bibitem{CHP} P. Cohadon, A. Heidmann, and M. Pinard, Phys. Rev. Lett. {\bf 83}  3174 (1999), Eprint: quant-ph/9903094.

\bibitem{Tittonen98} I. Tittonen, G. Breitenbach, T. Kalkbrenner,
T. M\"uller, R. Conradt, S. Schiller, E. Steinsland, N. Blanc and
N. F. de Rooij, to appear in Phys. Rev. A. {\bf 59}, 1038 (1999). 

\bibitem{adapt} H.M. Wiseman, Phys. Rev. Lett. {\bf 75}, 4587 (1995); H.M. Wiseman and R.B. Killip, Phys. Rev. A {\bf 57}, 2169 (1998); D. Berry, H.M. Wiseman and Z.X. Zhang, Phys. Rev. A {\bf 60}, 2458 (1999). 

\bibitem{vonN} J. von Neumann, {\em Tomsk Univ.\ Rev.}, {\bf 1}, 286 (1937), Reprinted in {\it John von Neumann: Collected Works\/}, vol. IV, (Macmillan, New York, 1962), edited by A. H. Taub.

\bibitem{FJ} C. Fuchs and K. Jacobs, `An information trade-off for finite strength quantum measurements', (in preparation).

\end{thebibliography}
\end{document}